\documentclass{moriond}

\def\Journal#1#2#3#4{{#1} {\bf #2}, #3 (#4)}

\def\be{\begin{equation}}
\def\ee{\end{equation}}
\def\bea{\begin{eqnarray}}
\def\eea{\end{eqnarray}}

\newcommand{\euler}{\mathrm{e}}                                
\usepackage{relsize,exscale}

\usepackage{ragged2e} 
\usepackage[most]{tcolorbox}

\tcbset{colback=yellow!10!white, colframe=blue!90!black, 
        highlight math style= {enhanced, 
            colframe=red,colback=red!10!white,boxsep=0pt}
        }
\usepackage{framed}
\usepackage[T1]{fontenc}
\usepackage[utf8]{inputenc}
\usepackage{geometry}
\usepackage{graphicx}
\usepackage{movie15}
\usepackage{sidecap}
\usepackage{color}
\definecolor{bole}{rgb}{0.47, 0.27, 0.23}
\usepackage{multicol}
\usepackage{caption}
\usepackage{color}
\usepackage{amsmath,color}

\usepackage{amsmath}
\usepackage{xcolor}

\definecolor{bole}{rgb}{0.47, 0.27, 0.23}                  
\definecolor{llgray}{gray}{0.90}                           
\definecolor{ddgray}{gray}{0.60}                           
\definecolor{dlgray}{gray}{0.40}                       
\definecolor{ldgray}{gray}{0.39}
\definecolor{vlgray}{gray}{0.98}
\usepackage{hyperref}
\hypersetup{colorlinks, linkcolor={red}}
\usepackage{stackengine,scalerel}
\newcommand\dAlaux{%
  \Shortstack{\rule{12pt}{.6pt}\\
    \rule{.6pt}{10pt}\kern10pt\rule{1.4pt}{10pt}\\
    \rule{12pt}{1.4pt}}%
}
\newcommand\dAl{%
  \setstackgap{S}{0pt}%
  \setstackEOL{\\}%
  \scalerel*{\kern1pt\dAlaux\kern1pt}{\Delta}%
}

\usepackage{ragged2e}

\usepackage{lipsum}
\usepackage{multicol}

\newcommand{\Photo}{}

\begin{document}
\vspace*{4cm}
\title{WAVE-FRONTS OF GRAVITATIONAL WAVES PARTIALLY TRAPPED IN ULTRA-COMPACT STARS\,\footnote{Based on an unpublished paper by Hor{\'a}k, Klimovi{\v c}ov{\'a} and Abramowicz \cite{hoklab}.}}
\author{Marek Abramowicz}
\address{Department of Physics, G{\"o}teborg University, SE-412-96 G{\"o}teborg, Sweden\\
Department of Physics, Silesian University in Opava, Czech Republic\\
N.Copernicus Astronomical Centre (Camk, PAN), Warsaw, Poland}
\author{Ji{\v r}{\'{\i}} Hor{\'a}k}
\address{Institute of Astronomy, Czech Academy of Sciences, Prague, Czech Republic}
\author{Kate{\v r}ina Klimovi{\v c}ov{\'a}}
\address{Department of Physics, Silesian University in Opava, Czech Republic}
\maketitle\abstracts
{
We dedicate this work to Dr Omer Blaes, professor of physics at UCSB, on the occasion of his sixtieth birthday. We have been collaborating now and then with Dr Blaes on problems involving oscillations, waves and stability\,\cite{abrblahor}. Happy birthday, Omer. Enjoy the analytic treatment of damping of the gravitational waves trapped inside ultra compact stars and its possible connection to Quantum Gravity in the context of the LIGO-Virgo efforts in accurately measuring ringdowns and echoes.
}
\section{Quantum Gravity is coming to town.}
\label{sec:experimental-QG}

There is a widespread hope that the recent advances in technology for observing the black hole high energy phenomena --- in particular the gravitational wave detections by LIGO-Virgo (LV) and the close-up imaging of black holes in M87* and SgrA* by the Event Horizon Telescope (EHT) --- could provide means of a direct check on some Quantum Gravity (QG) ideas. Cardoso and Pani in an excellent recent review\,\cite{carpa} elaborate on the consensus opinion that some of the features anticipated in observations by LV (e.g. the echoes  and/or the non-standard ringdown shape) and by EHT (e.g. ``the second ring'') may distinguish between the Kerr black hole and some of its alternatives --- e.g. strange stars, gravastars, wormholes, firewalls --- by {\it proving} that the objects observed are horizonless. 

However, there is a minority opposition to this, most eloquently expressed by the late Freeman Dyson\,\cite{dys1}$^{,\,}$\cite{dys2}, who points out that QG is meaningless if we have no apparatus able to detect a single graviton. Dyson formally proved that LV, even when updated, cannot for a fundamental reason detect a single graviton. In this paper we reply to this criticism by recalling an analogy with electromagnetic radiation. Physicists long before being able to detect single photons, had at hand proofs of the quantum nature of light, e.g. by observing the Fraunhofer lines. Here we suggest, although we do not prove, that in the {\it observed} ringdown phenomenon there could be detectable imprints of the fact that the energy of the gravitational radiation is carried in finite portions --- the gravitons.   

\section{The Teukolsky wave equation for the internal Schwarzschild metric in the optical geometry representation.}
\label{sec:schwarzschild-optical}
Let us rewrite the well-known Schwarzschild solution describing the vacuum metric outside a spherically symmetric, static body with mass $M$ (and gravitational radius $r_{\rm G} = GM/c^2$) in the ``optical geometry'' form:
\begin{eqnarray}
ds^2 &=& \left(1 - \frac{\mathlarger r_{\rm G}}{\mathlarger r}\right) \left[ dt^2 - 
 \left( 1 - \frac{\mathlarger r_{\rm G}}{\mathlarger r}\right)^{-2}dr^2 - \left(1 - \frac{\mathlarger r_{\rm G}}{\mathlarger r}\right)^{-1} r^2 (d\theta^2 + \sin^2\theta 
d\phi^2)\right] \\
&=& \euler^{-2\Phi} \left[ dt^2 - d{r_*}^2 - {\tilde r}^2
(d\theta^2 + \sin^2\theta d\phi^2) \right] = \euler^{-2\Phi} \left[ dt^2 - dh^2\right].
\label{eq:general-optical} 
\end{eqnarray}
The optical geometry appears in the square brackets --- it is conformal to the original Schwarzschild metric. The second line (\ref{eq:general-optical}) represents optical geometry for any static, space-time geometry in spherical coordinates. The 3-D ``optical space'' has the metric $-dh^2 = \gamma_{ik} dx^i dx^k$. 

In the Minkowski metric, $\Phi = 0$, and $r_* = {\tilde r} = r$, but in general, one has
\begin{equation}
\Phi = -\frac{1}{2}\ln g_{tt}, ~~r_* = {\mathlarger \int} {\mathlarger {\sqrt{-\frac{g_{rr}}{g_{tt}}}}}dr, ~~ {\tilde r} = \sqrt{-\frac{g_{\phi \phi}}{g_{tt}}}.
\label{eq:radii} 
\end{equation}
Along a null geodesic, i.e. along a light or gravitational wave-front trajectory $ds=0$.
From the Fermat principle, $\delta\int dt = 0$ along light trajectories in static space-times. 
Thus, from (\ref{eq:general-optical}) one concludes that $\delta\int dh = 0$ along light trajectories: they are geodesic lines also in the 3-D optical space\,\cite{abcala}. 

The optical geometry corresponding to the interior of a constant density Schwarzschild star is 
spherical and isometric with spatial sections of the static Einstein Universe with the curvature scalar ${\cal R} = 6/{\tilde a}^2=$\,const. The 3-D optical space corresponding to the internal Schwarzschild solution has the metric,
\begin{equation}
dh^2 = \left[ dr_\ast^2 + {\tilde r}^2\left( d\theta^2  + \sin^2\theta \,d\phi^2 \right)\right], 
~~{\tilde r} = {\tilde a}\sin(r_\ast/{\tilde a}).
\label{eq:optical-space}
\end{equation}
The spherical geometry of the 2-D equatorial plane $\theta=\pi/2,\, d\theta=0$ is shown as an embedding diagram in Figure \ref{fig:star}. The interior (indicated by the color) and the exterior Schwarzschild solution, join at the radius of the star $r=R<(3/2)r_{\rm G}$. Note that the radius ${\tilde a}$ of the spherical bulge equals (we use geometrical units with $c=1=G$),
\begin{equation}
{\tilde a} = \frac{R}{2}\left( \frac{R}{M}\right)^{1/2} \left( 1 - \frac{9}{4} \frac{M}{R} \right)^{-1/2}
\label{eq:a-tylde}
\end{equation} 
where $R$ is the radius of the star and $M$ is its mass. For compactness $M/R = 4/9$ the radius of the bulge tends to infinity. This is the maximal compactness a non-singular Schwarzschild star could have --- the Buchdahl-Bondi limit\,\cite{buch}.   

Let the wave function be described in the Minkowski space-time by the standard formula, 
\begin{equation}
\Psi(t, r, \theta, \phi) =
\mathop{\mathlarger{ \sum_{\ell=0}^{\infty}}}
\mathop{\mathlarger{\sum_{m=-\ell}^{\ell}}}   
\left\{ 
W_{ \ell m}(r) P^m_\ell (\cos \theta) \, \rm{e}^{-\rm{i} \omega t}\, \rm{e}^{-\rm{i} m \phi} 
\right\}.
\end{equation}
$P^m_\ell (\cos \theta)$ are the Legendre polynomials. After the standard separation of variables, the radial part of the wave equation in the axially symmetric case $\partial_\phi = 0,\,m=0$, takes the familiar form:
\begin{equation}
\mbox{Minkowski:}~~~ \dAl_{\rm r} W
= \left[ 
\, \omega^2 - 
 \frac{\mathlarger{\partial^2}}
 {\mathlarger{\partial r^2 }} 
+  
V(r)                                                                     
\right] W = 0, ~~\mbox{where}~~V(r)=\frac{\mathlarger \ell (1+\ell)}{\mathlarger r^2}.
\label{eq:radial-wave-Minkowski}
\end{equation}
Using the optical geometry reasoning based on the spirit of equation (\ref{eq:general-optical}), one may immediately guess the correct form of the wave equation (``the Teukolsky equation'') in a general static space-time, just by removing the Minkowski degeneracy $r = r_* = {\tilde r}$. 
\begin{equation}
\mbox{General:}~~~ \dAl_{\rm r} W
= \left[ 
\, \omega^2 - 
 \frac{\mathlarger{\partial^2}}
 {\mathlarger{\partial r_*^2 }} 
+  
V(r)                                                                     
\right] W = 0, ~~\mbox{where}~~V(r)= \frac{\mathlarger \ell (1+\ell)}{\mathlarger {\tilde r}^2} + \delta V(r).
\label{eq:radial-wave-Teukolsky}
\end{equation}
The frequency of the wave is a complex quantity, with ${\rm Re}(\omega)= \sigma$ being the frequency of the quasi-normal modes, and ${\rm Im}(\omega)= 1/t_0$ being the inverse of a characteristic damping time of these modes, i.e. of the ``ringdown'' in the LV language. 

\section{Formal solution of the Teukolsky equation.}
\label{sec:teukolsky-formal}

It would be convenient to introduce new variables (see Figure \ref{fig:star}), $\chi \equiv r_\ast/\tilde{a}, ~ \kappa \equiv \tilde{a}\omega$, and write the Teukolsky equation (\ref{eq:radial-wave-Teukolsky}) in a dimensionless form,
\begin{equation}
\frac{\mathlarger{ d^2 W}}{\mathlarger{ d\chi^2}} + \left[\kappa^2 - \tilde{a}^2\left(V_\ell + \delta V\right)\right]W = 0, ~~ \mbox{where} ~~ V_\ell \equiv \frac{\ell(\ell+1)}{{\tilde r^2}} \gg \delta V
\label{eq:Teukolsky-dimensionless}
\end{equation}
We solve the Teukolsky equation separately for the exterior and interior Schwarzschild solutions, adopting the regularity condition at the center and no ingoing wave at infinity,
\begin{equation}
W^\prime(0) = 0, \quad \left[W^\prime - \mathrm{i}\kappa W\right]_{\chi\rightarrow\infty} = 0
\end{equation} 
The exterior solution is given by the WKBJ approximation, 
\begin{equation}
W_\mathrm{ext} = \left(\frac{\Theta}{\gamma}\right)^{1/4}\left[\mathrm{Ai}(\Theta) - \mathrm{i}\mathrm{Bi}(\Theta)\right],
\end{equation}
with $\mathrm{Ai}(\Theta)$ and  $\mathrm{Bi}(\Theta)$ being the Airy functions,
and 
\begin{equation}
\Theta \equiv \left[\frac{3}{2}\mathlarger{\int}_{r(\chi)}^{r_\mathrm{t}}\gamma(r)\left(\frac{d\chi}{dr}\right)dr\right]^{2/3}, ~~\gamma(r) \equiv \sqrt{\tilde{a}^2 V_\mathrm{ext}(r) - \kappa^2}. 
\end{equation}
The interior solution is given in terms of the associated Legendre functions: 
\begin{equation}
W_\mathrm{int}(\chi) = (\sin\chi)^{1/2} Q^\mu_\nu(\cos\chi), ~~~~\mu=\ell+1/2, ~~\nu=\kappa-1/2
\end{equation} 
Matching the interior and exterior solutions at the surface of the star $r=R$ reveals the eigenvalue nature of the problem. The matching is possible only for the eigenfrequencies $\kappa$ characterized by
\begin{equation}
\mathrm{Re}(\kappa) = n, ~~~~
\mathrm{Im}\left(\kappa\right) = - A_\ell \frac{\mathlarger{n^{2\lambda}(\ell+n)!}}{\mathlarger{(n-\ell-1)!}} 
\times	\exp\left(\frac{32}{27}\pi n x^{1/2}\right)x^{\lambda+\ell+1/2}.
\label{eq:formal-frequencies}
\end{equation}
Here $\lambda^2 = \ell(\ell + 1)$, $A_2 = 1.165 \times 10^{-2}$ and $x = 1 - 9M/4R$.


\section{Optical geometry solution for ${\rm Re}(\kappa)$ and ${\rm Im}(\kappa)$.}
\label{sec:teukolsky-optical}
%
\begin{multicols}{3}
\begin{center}
\includegraphics[width=0.90\columnwidth]{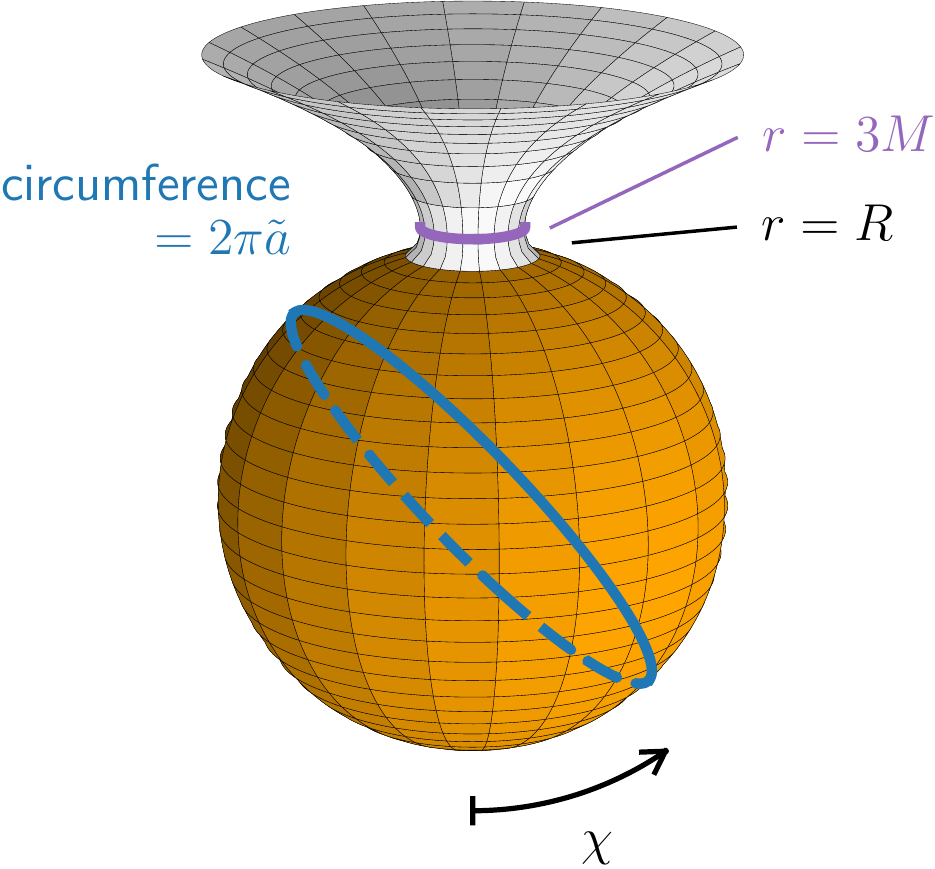}\label{fig:star}
\end{center}
\noindent {\footnotesize Figure \ref{fig:star}: The Schwarzschild optical space
($\theta=\pi/2$ cut). The interior solution is indicated by the color.
}
\vfill
\columnbreak                                                        
\noindent {\footnotesize  $\chi$ is the rescaled proper distance from the center of the star, and $\tilde a$ is
the circumferential radius of great circles at the spherical bulge of the stellar interior: the great circles are
geodesic lines in the metric (\ref{eq:optical-space}), therefore they are light trajectories in the optical space. It was noticed\,\cite{abrandbru} that a standing wave that has $n$ nodes along a great circle, obviously has frequency
\vskip0.1truecm \noindent 
\begin{center}
\hskip1.0truecm $\kappa= n$ \hfill (a)
\end{center}
\vskip0.1truecm \noindent 
This agrees with the formally derived equation (\ref{eq:formal-frequencies}) for Re($\kappa$). }
\vfill
\columnbreak                                                        
\noindent{\footnotesize
One may estimate the mode decay time $t_0$ by considering the black body radiation with a uniform energy density $\varepsilon$ enclosed in a container. The radiation flux that leaks from a small hole in the container is $f=4\varepsilon/c$. Therefore,
$t_0 = {\varepsilon \cal{V}}/{f{{\cal A}}}$, 
where ${\cal V}(R)$ is the volume of the container and ${\cal A}(R)$ is the surface area of the hole. Using the explicit form of these two functions that are known in the Schwarzschild internal solution, one arrives at (for $\ell=2$):
\vskip0.1truecm \noindent
\begin{center} 
\hskip1.0truecm $t_0 ={\rm const}\, x^{-3/2}$ \hfill (b)
\end{center}
}
\vfill
\end{multicols}

\section{Discussion}
\label{sec:discussion}
%
\begin{multicols}{3}
\vskip-0.1truecm
\begin{center}
\includegraphics[width=0.95\columnwidth]{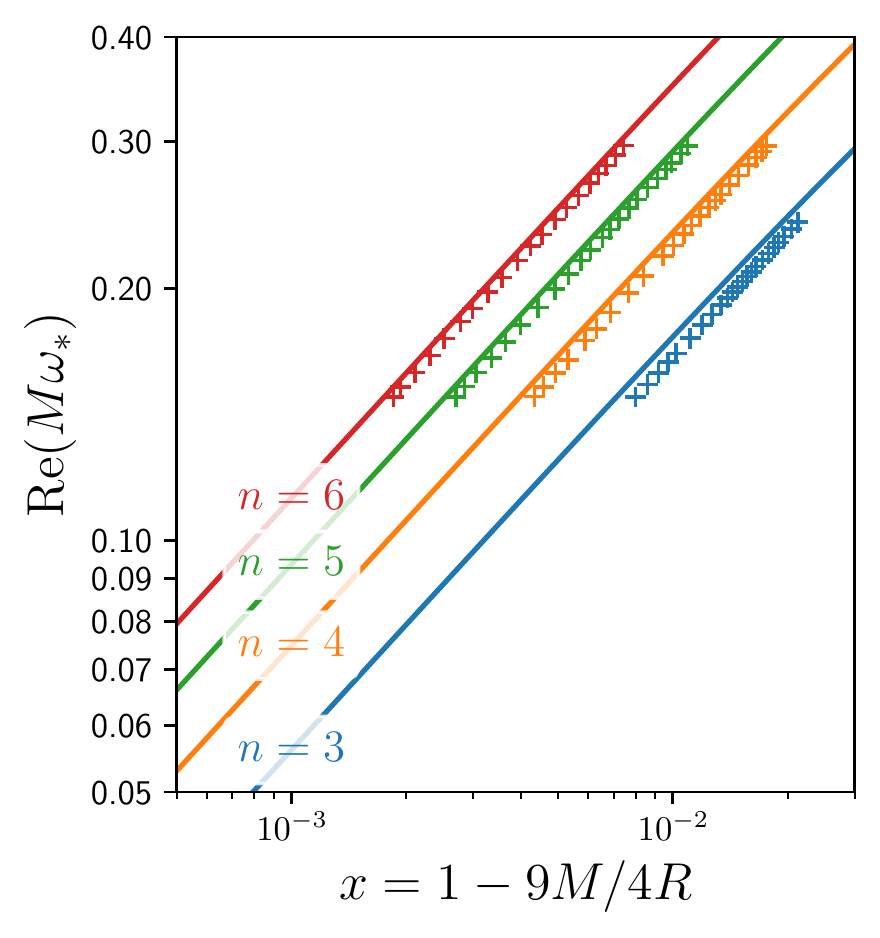}\label{fig:real}
\end{center}
\noindent {\footnotesize Figure \ref{fig:imaginary}: Solid lines: equations (\ref{eq:formal-frequencies}) and (a). Points: numerical\,\cite{andkoj}. 
}
\vfill
\columnbreak                                                        
\begin{center}
\includegraphics[width=0.98\columnwidth]{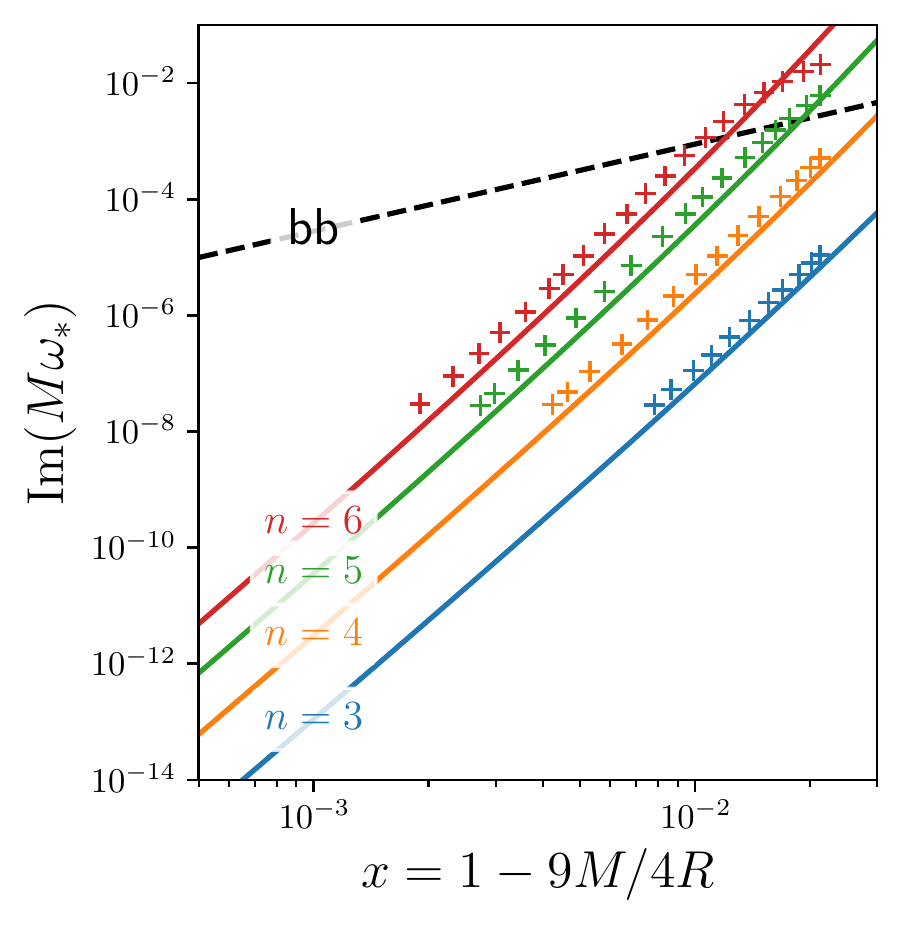}\label{fig:imaginary}
\end{center}
\vskip0.05truecm
\noindent {\footnotesize Solid lines: (\ref{eq:formal-frequencies}). Dashed line bb: eq. (b). They do {\it not} coincide. 
}
\vfill
\columnbreak                                                        
\noindent{\footnotesize
Our formal calculation of $t_0$ is based on the classical, non-quantum, Teukolski equation, while the bb line in the Figure is calculated from the quantum black body EM Planck formula: radiation in the container consists of individual photons and is radiated away photon by photon. If this makes any practical difference, the ringdown shape, {\it observed sufficiently accurately}, should therefore be affected by the existence of gravitons!
}
\vfill
\end{multicols}

\section{Conclusions and plans}
\label{sec:conclusions}

{\footnotesize 
Discrete orbital structures (i.e. ones described by natural numbers)  are typical of many aspects of classical celestial mechanics: well-known examples are Newton's orbital resonances or Einstein's holonomy invariance of parallel transport\,\cite{rotell}$^,$\,\cite{abraalme}, and the kilohertz quasi-periodic oscillations\,\cite{torabr}$^-$\,\cite{Frastra}. In all of these situations the gravity has no discrete aspects, only the orbits have them. All of these effects are observed.

In this paper we described the discrete features of the ``ringdowns'' i.e. damped gravitational wave-fronts: (a)~classically, by solving the Teukolsky equation and (b)~quantum-mechanically, by employing a supposed graviton-photon analogy, based on the black body formulae. These two descriptions give different results. We argue that in a more sophisticated theoretical analysis of the ringdowns than presented here, the discrete nature will  mature into a kind of the ``Planck formula for gravitational radiation''. Because the ringdowns {\it are observed} by the LV interferometers, this will be a sure step towards experimental quantum gravity. 

We are now working on a few follow-ups: (i) A ``topology catalogue'' of ringdowns, reflections and echoes for ``topology different'' ultra-compact objects. (ii) A non-linear Teukolsky equation and possible GW gravitational resonances. (iii) Application of the Rayleigh-Jeans law into wave turbulence: different wave modes exchange
their energy (like nonlinear harmonic oscillators) to reach equipartition. One may apply the same to GWs and perhaps repeat the RJ derivation with a Planck-like ansatz.
} 

\section*{Acknowledgments}

{\footnotesize This work was supported by the Czech Science Foundation grant EXPRO 21-06825X.

\end{document}